# Generic Microstrip R&D Topics at SCIPP: Longitudinal Charge Division and Length Limitations for Long Strips

Jerome K. Carman, Vitaliy Fadeyev, Khilesh Mistry, Bruce A. Schumm, Edwin Spencer, Aaron Taylor and Max Wilder

Santa Cruz Institute for Particle Physics and the University of California at Santa Cruz Department of Physics
1156 High Street, Santa Cruz, CA 95064 – USA

We discuss results on the use of charge division to estimate the longitudinal position of charge deposition in silicon microstrip sensors, and on the intrinsic noise limitation of microstrip sensors in the limit of long, narrow strips. We find a resolution of ±6% of the length of the sensor for a 10cm-long sensor. We also find that network effects significantly reduce sensor readout noise relative to naïve expectations that arise from treating the detector resistance and capacitance as single discrete electronic components.

With its need to instrument large area with ultra-precise position measurement, designing a silicon microstrip tracker for use in a Linear Collider Detector presents a number of challenges. While no longitudinal information is required of the central tracking system to achieve the momentum resolution necessary to exploit Linear Collider Physics, studies [1] have shown that longitudinal information with a resolution of 1cm or better can aid the unraveling of charged-particle trajectories in the high-energy jets that are produced in Linear Collider collisions. In addition, to avoid the complexity, the material and thermal burden, and to reduce the mechanical shock associated with power cycling great numbers of electronic channels in a large magnetic field, it may prove advantageous to reduce the number of read-out channels by reading out daisy-chained "ladders" of strip sensors with a length approaching 1m. To this end, in this proceeding we explore the use of charge division to estimate the longitudinal coordinate of charge deposition within silicon microstrip sensors, and the intrinsic noise limitations associated with reading out long, narrow sensor strips. These studies shed light on different optimization strategies for an all-silicon central tracker.

## 1 Longitudinal charge division

If electronic charge or current is read out from two opposing ends of a microstrip sensor, and the impedance of the read-out circuitry is small compared to the resistance of the sensor electrode, it is possible to measure the position of a localized deposition of charge within the sensor by comparing the relative amplitudes of the signals amplified by the two opposing read-out circuits. In lieu of sensors with resistive electrodes, a printed circuit (PC) board was designed with electronic network characteristics similar to those of a microstrip sensor. The electrode was divided into ten segments, with the total electrode resistance set at 600 kΩ by the value of discrete resistors placed between the nine nodes between the segments. The total capacitance to ground of the electrode, due to parasitic coupling, was measured to be 13 pC, close to that expected for a 10 cm long strip. Since read-out noise is dominated by the capacitance and Johnson noise of the electrode, off-the-shelf electronics (the Texas Instruments OPA657 low-noise FET input operational amplifier and the Analog Devices model ADA4851 video op-amp) were used to develop an amplification network with a tunable shaping time. Both ends of the resistive electrode were instrumented with read-out electronics. A probe was used to inject known values of charge at any of the nine nodes of the electrode, allowing response to be measured as a function of position along the electrode.

A PSpice [2] simulation was used to optimize the shaping time of the electronic read-out, chosen to provide a balance between signal-to-noise and the linearity of response with respect to the node at which the charge was injected. After tuning the shaping time of the read-out system to this optimal value, good agreement between observation and simulation was observed, for injection of charge at all nodes, for both the temporal response (Fig. 1) and the signal-to-noise of the read-out.

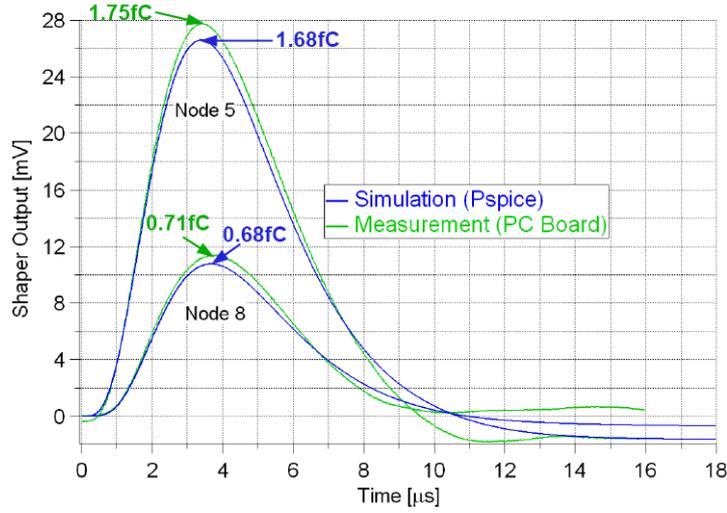

Figure 1: Predicted and measured excitation curve for the charge division readout, for nodes at the half-way and 80% points of the network.

In terms of the charges $Q_R$, $Q_L$ measured on the right and left side of the network, respectively, and the observed input-referenced noise $\sigma$, the fractional longitudinal resolution $\sigma_P$ is given by

$$\sigma_P = \left(\frac{1}{(1+\alpha)^2}\right)\sigma_\alpha$$

with $\alpha = Q_R/Q_L$, and

$$\sigma_\alpha = (\alpha)\left\{\left(\frac{\sigma_R}{Q_R}\right)^2 + \left(\frac{\sigma_L}{Q_L}\right)^2 - 2\rho\left(\frac{\sigma_R}{Q_R}\right)\left(\frac{\sigma_L}{Q_L}\right)\right\}^{\frac{1}{2}},$$

where $\rho$ is the correlation coefficient between the left- and right-hand amplifier noise values, measured to be $\rho = -0.61$ from simultaneous acquisitions of the left- and right-hand response to a fixed input charge. Assuming a mean charge deposition of 3.7 fC for a minimum-ionizing particle, this leads to a fractional resolution of 6.0%, essentially independent of the longitudinal position of the charge deposition. For a 10cm sensor, this would provide a longitudinal coordinate to an accuracy of 0.6cm, well within the range of interest for the separation of individual charged tracks in collimated Linear Collider jets. A more detailed accounting of the project and its results are to be found in [3].

## 2 Readout noise in the long, narrow strip limit

One approach to reducing the material and power budget, and limiting electromagnetic-induced vibration from power-pulsing, is to build Linear Collider central tracking from long (~1m) microstrip sensor ladders, limiting the number of electronic channels needed to read out the tracker. Since the single-hit position measurement accuracy would need to be maintained, the sensor pitch would need to remain small, which would push the electronic characteristics of the strip into a high-capacitance (~100 pF), moderate-resistance (~1kΩ) regime – a regime that has yet to be explored or implemented with any existing microstrip system.

Based on an assumption of lumped capacitance and resistance, Spieler [4] has characterized electrode readout noise Q in equivalent electrons as

$$Q^2 = F_i \tau \left( 2eI_d + \frac{4kT}{R_B} + i_{na}^2 \right) + \frac{F_v C^2}{\tau} (4kTR_s + e_{na}^2) + 4F_v A_f C^2$$

where $\tau$ is the amplifier-chain shaping time, C the strip capacitance, $R_S$ the strip resistance, $R_B$ the parallel (biasing) resistance, and $I_d$ the detector leakage current. $i_{na}$ and $e_{na}$ are the amplifier current and voltage noise, respectively. The "shape factors" $F_i$ and $F_v$ depend on moments of the amplifier response function, and are of order 1; $A_f$ is set by bulk defect noise and is usually not appreciable. Although this expression has been often used to rough out potential readout designs, realistic sensor strips are not represented by lumped electronic values, but are continuously distributed RC networks, which could potentially behave quite different to their lumped analogs. Thus, we have undertaken an exploration of realistic readout noise in the long, narrow strip limit.

To measure readout noise as a function of distributed capacitance and resistance, we connected the low-noise LSTFE-I front-end amplifier to a single SiD prototype sensor (one of the "charge division" sensors for which metallization was mistakenly applied above the implant), which had a distributed per-strip capacitance of 5.2 pF and resistance of 287 Ω. This allowed for the attachment of successive strips by means of an up-and-back "snake" pattern. In this way, a total of up to 13 strips were read out, with a total distributed capacitance of 67.6 pF and resistance of 3730 Ω.

Fig. 2 shows the measured readout noise as a function of the number of read-out strips, compared to two models: the lumped model of Ref. [4], and a Spice [2] simulation. For the simulation, the measured noise performance of the LSTFE-I has been introduced via an ideal noise source at the preamplifier input that provides a tunable amount of voltage-referenced noise but left the network characteristics unchanged. Good agreement is observed between the PSpice model and the measurements, once the LSTFE-I noise contributions have been accounted for. It is observed that: a) network effects appear to significantly mitigate the readout noise for long, narrow strips; and b) the noise can be further reduced by reading the chain out from the center rather than from its ends. This is good news, in that it improves the viability of long-ladder solutions for the ILC. These results are in preparation for publication.

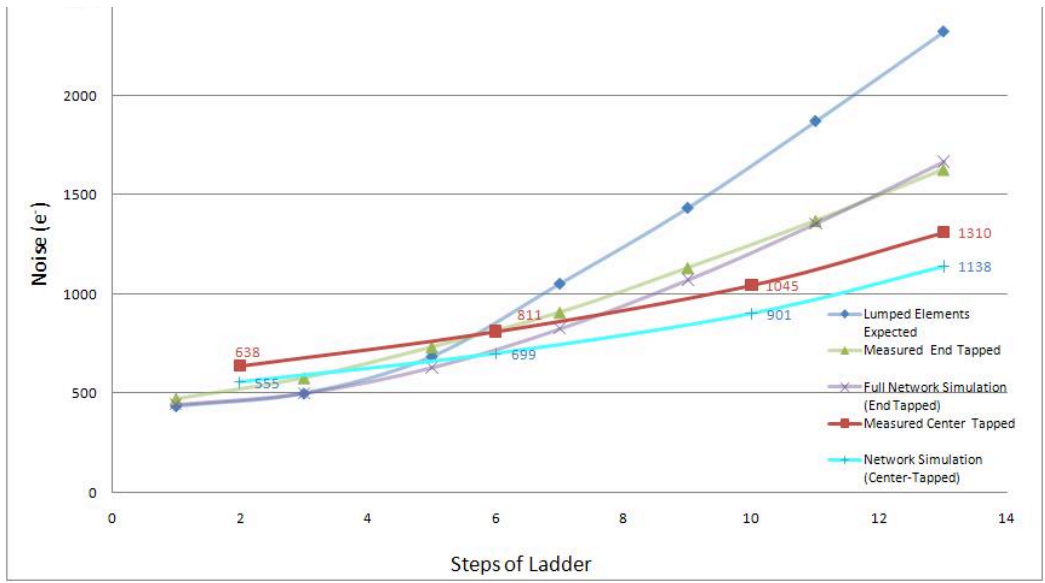

Figure 2: Measured noise as a function of load for end-readout (green) and center- readout (red), compared to the full (brown, blue-green) and lumped (dark blue) network models.

## Acknowledgments

The authors acknowledge and are grateful for the support provided by the United States Department of Energy, under contracts DE-SC0001476 (subaward 234171B) and DE-FG02-04ER41286, which enabled the successful execution of this work.

## References

[1] C. Meyer, T. Rice, B. A. Schumm and L. Stevens, *Simulation of an All-Silicon Tracker*, arXiv:0709.0758 [physics.ins-det] (Sept 2007).

[2] L. W. Nagel, D. O. Pederson, SPICE (Simulation Pprogram with Integrated Circuit Emphasis), EECS Department, University of California, Berkeley Technical Report No.UCB/ERLM382, (April 1973).

[3] J. K. Carman et al., Nuclear Instruments and Methods in Physics Research A 646 (2011) 118–125.

[4] H. Spieler, *Semiconductor Detector Systems*, Oxford Univertsity Press, 2005